\documentclass{JHEP3}
\usepackage{epsfig}
\preprint{UT-STPD-3/02\\ IPPP/02/37\\ DCPT/02/74}
\title{New shifted hybrid inflation}
\author{R. Jeannerot\\ INFN-SISSA/ISAS, via
Beirut 2-4, Trieste 34013, Italy\\ E-mail:
\email{rachelj@sissa.it}}
\author{S. Khalil\\ IPPP, Physics Department,
Durham University, Durham DH1 3LE, U.K., and\\
Faculty of Science, Ain Shams University,
Cairo 11566, Egypt\\
E-mail: \email{shaaban.khalil@durham.ac.uk}}
\author{G. Lazarides\\ Physics Division,
School of Technology, Aristotle University
of Thessaloniki, Thessaloniki 54124, Greece\\
E-mail: \email{lazaride@eng.auth.gr}}

\abstract{A new shifted hybrid inflationary
scenario is introduced which, in contrast
to the older one, relies only on
renormalizable superpotential terms. This
scenario is automatically realized in a
concrete extension of the `minimal'
supersymmetric Pati-Salam model which
naturally leads to a moderate violation of
Yukawa unification so that, for $\mu>0$,
the predicted $b$-quark mass is acceptable
even with universal boundary conditions. It
is shown that this extended model possesses
a classically flat `shifted' trajectory
which acquires a slope via one-loop
radiative corrections and can be used as
inflationary path. The constraints from the
cosmic background explorer can be met with
natural values of the relevant parameters.
Also, there is no disastrous production of
magnetic monopoles after inflation since
the Pati-Salam gauge group is already
broken on the `shifted' path. The relevant
part of inflation takes place at values of
the inflaton field which are not much
smaller than the `reduced' Planck scale and,
thus, supergravity corrections could easily
invalidate inflation. It is, however, shown
that inflation can be kept intact provided
that an extra gauge singlet with a
superheavy vacuum expectation value, which
originates from D-terms, is introduced and
a specific form of the K\"{a}hler potential
is used. Moreover, it is found that,
although the supergravity corrections are
sizable, the constraints from the cosmic
background explorer can again be met by
readjusting the values of the parameters
which were obtained with global
supersymmetry.}

\keywords{Cosmology of Theories beyond the
SM, Cosmological Phase Transitions, Physics
of the Early Universe}

\begin{document}

\section{Introduction}
\label{sec:intro}

\par
The recent measurements \cite{peaks} of the three
acoustic peaks of the angular power spectrum of
the cosmic microwave background radiation (CMBR)
have strongly favored the idea of inflation
\cite{inflation} (for a review see e.g.,
Ref.\cite{lectures}). Moreover, inflation is now
considered as the most likely explanation of the
origin of structure formation
in the universe. Therefore, it is important to
construct viable models of inflation which are
consistent with all particle physics and
cosmological requirements. Undoubtedly, one of
the most promising inflationary scenarios is
hybrid inflation \cite{linde}. It uses two real
scalar fields: a gauge non-singlet which
provides the vacuum energy density needed for
driving inflation and a gauge singlet which is
the slowly varying field during inflation.
Hybrid inflation is \cite{cop} naturally
incorporated in supersymmetric (SUSY) grand
unified theory (GUT) models based on gauge
groups with rank greater than or equal to five
(for specific successful models see e.g.,
Ref.\cite{models}).

\par
One of the most attractive rank five gauge groups
is certainly the Pati-Salam (PS) group $G_{PS}=
SU(4)_c\times SU(2)_L\times SU(2)_R$ \cite{pati}.
This is the simplest GUT gauge group which can
lead \cite{hw} to Yukawa unification \cite{als}.
It can also generate seesaw masses for the light
neutrinos and has \cite{shafi} many other
interesting phenomenological implications.
Moreover, SUSY PS GUT models are motivated
\cite{tye} (see also Ref.\cite{kane}) from the
recent D-brane setups and can also arise
\cite{leontaris} from the standard weakly coupled
heterotic string.

\par
A characteristic feature of standard SUSY hybrid
inflation, which is based on a renormalizable
superpotential, is that the spontaneous breaking
of the GUT gauge symmetry takes place at the end
of inflation and thus topological defects may be
copiously formed \cite{smooth}. In particular,
the spontaneous breaking of $G_{PS}$ to the
standard model (SM) gauge group $G_S$ at the end
of standard SUSY hybrid inflation leads
\cite{shift} to the overproduction of topologically
stable magnetic monopoles, which carry \cite{magg}
two units of Dirac magnetic charge. So a
cosmological disaster is encountered.

\par
A way out of this potential catastrophe in hybrid
inflationary models is to include into the standard
superpotential for hybrid inflation the leading
non-renormalizable term \cite{shift} (see also
Ref.\cite{smooth}; for a summary see
Ref.\cite{talks}). It was observed \cite{shift}
that this term cannot be excluded by any symmetry
and can be comparable to the trilinear term of
the standard superpotential. The coexistence of
both these terms leads \cite{shift} to the
appearance of a new `shifted' classically flat
valley of local minima where the GUT gauge
symmetry is broken. This valley acquires a slope
at the one-loop level and can be used \cite{shift}
as an alternative inflationary path. In this
scenario, which is known as shifted hybrid
inflation, there is no formation of topological
defects at the end of inflation and hence the
potential monopole problem is avoided. This is
crucial for the compatibility of the SUSY PS
model with hybrid inflation since this model
predicts the existence of magnetic monopoles.

\par
It would be desirable to solve the (potential)
magnetic monopole problem of hybrid inflation in
SUSY GUTs with the GUT gauge group broken
directly to $G_S$ without relying on the presence
of non-renormalizable superpotential terms. (The
monopole problem could also be solved by employing
\cite{twostep} an intermediate symmetry breaking
scale or by other mechanisms (see e.g.,
Ref.\cite{fate}).) We shall, thus,
examine the possibility of having shifted hybrid
inflation in a SUSY PS GUT by using only
renormalizable interactions and with $G_{PS}$
still broken to $G_S$ in a single step.

\par
In this paper, we show that a new version of
shifted hybrid inflation can take place in the
SUSY PS model without invoking any
non-renormalizable superpotential terms provided
that we supplement the model with a new
conjugate pair of superfields, $\phi$,
$\bar{\phi}$, belonging to the representation
(15,1,3) of $G_{PS}$. These fields lead to three
new renormalizable terms in the part of the
superpotential which is relevant for inflation.

\par
This extension of the SUSY PS model is also
motivated \cite{quasi} by the requirement that
Yukawa unification is moderately violated so that,
for $\mu>0$, the predicted bottom quark mass
resides within the experimentally allowed range
even with universal boundary conditions. It is
well-known \cite{hall} that, in models with Yukawa
unification (or with large $\tan\beta$ in general),
the $b$-quark mass receives large SUSY corrections
which, for $\mu>0$, lead to unacceptably large
values of $m_b$. The requirement that the SUSY
correction to $m_b$ is not excessive imposes
additional constraints on the parameter space
and non-universal soft SUSY breaking terms as
well as some deviation from the minimal
K\"{a}hler potential must be considered
\cite{shafi,kobayashi}. Another possible solution
to this problem is to assume a small violation of
Yukawa unification, which can be accommodated in
the SUSY PS model by introducing new Higgs
superfields as emphasized in Ref.\cite{quasi}. It
was shown that, in this case, one can get
satisfactory values of $m_b$ even with universal
boundary conditions and in accord with all other
phenomenological and cosmological requirements.

\par
This paper is organized as follows. In section
\ref{sec:model}, we consider the extended SUSY
PS model and show that it possesses a `shifted'
classically flat direction. We then construct
the mass spectrum on this trajectory and
calculate the one-loop radiative corrections.
We show that successful new shifted hybrid
inflation can take place along this path where
$G_{PS}$ is broken to $G_S$. Thus, the monopole
problem is avoided. In section \ref{sec:sugra},
we study the effect of supergravity (SUGRA) on
inflation which, as it turns out, takes place at
values of the inflaton field which are close to
the `reduced' Planck scale. We show that a
mechanism \cite{panag}
utilizing a specific K\"{a}hler potential can be
used so that the SUGRA corrections do not
invalidate inflation. However, the results which
were obtained with global SUSY must now be
readjusted. Our conclusions are summarized in
section \ref{sec:conclusion}.

\section{New shifted hybrid inflation}
\label{sec:model}

\par
The SUSY PS model of Ref.\cite{shift} which leads
to shifted hybrid inflation has been extended in
Ref.\cite{quasi} in order to allow a moderate
violation of the `asymptotic' Yukawa unification
so that, for $\mu>0$, an acceptable value of the
$b$-quark mass is obtained even with universal
boundary conditions. We consider this extended
model as the basis of our discussion here. The
breaking of $G_{PS}$ to $G_S$ is achieved by the
superheavy vacuum expectation values (VEVs) ($=
M_{GUT}\approx 2.86\times 10^{16}~{\rm GeV}$,
the SUSY GUT scale) of the right handed neutrino
type components $\nu_H^c$, $\bar{\nu}_H^c$
of a conjugate pair of Higgs superfields
$H^c=(\bar{4},1,2)$, $\bar{H}^c=(4,1,2)$. The
model also contains a gauge singlet $S$ which
triggers the breaking of $G_{PS}$. Finally, in
order to have Yukawa unification violated by an
amount which is adequate for $\mu>0$, a new
conjugate pair of superfields $\phi$,
$\bar{\phi}$ belonging to the (15,1,3)
representation of $G_{PS}$ is included. For
details on the full field content and
superpotential, the global symmetries, the charge
assignments, and the phenomenological and
cosmological properties of this model, the reader
is referred to Ref.\cite{quasi} (and \cite{shift}).

\par
The superpotential terms which are relevant for
inflation are all renormalizable and are given
by
\begin{equation}
W=\kappa S(H^c\bar{H}^c-M^2)-\beta S\phi^2+
m\phi\bar{\phi}+\lambda\bar{\phi}H^c\bar{H}^c,
\label{superpotential}
\end{equation}
where $M$ and $m$ are superheavy masses of the
order of $M_{GUT}$, and $\kappa$, $\beta$ and
$\lambda$ are dimensionless coupling constants.
These parameters are normalized so that they
correspond to the couplings between the SM
singlet components of the superfields. We can
take $M,~m,~\kappa,~\lambda>0$ by field
redefinitions. For simplicity, we also take
$\beta>0$, although it can be generally complex.

\par
The scalar potential obtained from $W$ is given
by
\begin{eqnarray}
V=\left\vert\kappa(H^c\bar{H}^c-M^2)-\beta\phi^2
\right\vert^2+\left\vert 2\beta S\phi-m\bar{\phi}
\right\vert^2+\left\vert m\phi+\lambda H^c
\bar{H}^c\right\vert^2\nonumber \\
+\left\vert\kappa S+\lambda\bar{\phi}
\right\vert^2\left(\vert H^c\vert^2+\vert\bar{H}^c
\vert^2\right)+{\rm D-terms},~~~~~~~~~~~~~~~~
\label{potential}
\end{eqnarray}
where the complex scalar fields which belong to
the SM singlet components of the superfields are
denoted by the same symbols as the corresponding
superfields. As usual, the vanishing of the
D-terms yields $\bar{H}^c\,^{*}=e^{i\vartheta}H^c$
($H^c$, $\bar{H}^c$ lie in the $\nu^c_H$,
$\bar{\nu}^c_H$ direction). We restrict ourselves
to the direction with $\vartheta=0$ which contains
the `shifted' inflationary path and the SUSY
vacua (see below). Performing an appropriate
global transformation, we can bring the complex
scalar field $S$ to the positive real axis. Also,
by a gauge transformation, the fields $H^c$,
$\bar{H}^c$ can be made positive.

\par
From the potential in Eq.(\ref{potential}), we
find that the SUSY vacuum lies at
\begin{equation}
\frac{H^c\bar{H}^c}{M^2}\equiv\left(\frac{v_0}
{M}\right)^2=\frac{1}{2\xi}\left(1-(1-4\xi)
^{\frac{1}{2}}\right),~~S=0,
~~\frac{\phi}{M}=-\frac{\kappa^{\frac{1}{2}}\xi
^{\frac{1}{2}}}{\beta^{\frac{1}{2}}}\left(
\frac{v_0}{M}\right)^2,~~\bar{\phi}=0,
\label{vacuum}
\end{equation}
where $\xi=\beta\lambda^2M^2/\kappa m^2<1/4$. Here,
we chose the vacuum with the smallest $v_0~(>0)$
for the same reasons as in Ref.\cite{shift}.
The potential possesses the trivial flat direction
at $H^c=\bar{H}^c=\phi=\bar{\phi}=0$ with
$V=\kappa^2M^4$. It also possesses a `shifted' flat
direction at
\begin{equation}
\frac{H^c\bar{H}^c}{M^2}\equiv\left(\frac{v}{M}
\right)^2=\frac{2\kappa^2(\frac{1}{4\xi}+1)+
\frac{\lambda^2}{\xi}}{2(\kappa^2+\lambda^2)},
~~S>0,~~\frac{\phi}{M}=-\frac{\kappa^{\frac{1}{2}}}
{2\beta^{\frac{1}{2}}\xi^{\frac{1}{2}}},
~~\bar{\phi}=-\frac{\kappa}{\lambda}S
\label{trajectory}
\end{equation}
with
\begin{equation}
\frac{V_0}{M^4}=\frac{\kappa^2\lambda^2}{\kappa^2+
\lambda^2}\left(\frac{1}{4\xi}-1\right)^2,
\label{V0}
\end{equation}
which can be used as inflationary path. As in the
case of the shifted hybrid inflationary model of
Ref.\cite{shift}, which is based on non-renormalizable
superpotential terms, the constant classical energy
density on the `shifted' path breaks SUSY, while the
constant non-zero values of $H^c$, $\bar{H}^c$ break
the GUT gauge symmetry. The SUSY breaking implies the
existence of one-loop radiative corrections which
lift the classical flatness of this path yielding the
necessary inclination for driving the inflaton
towards the SUSY vacuum.

\par
The one-loop radiative corrections to the potential
along the `shifted' inflationary trajectory are
calculated by using the Coleman-Weinberg formula
\cite{cw}:
\begin{equation}
\Delta V=\frac{1}{64\pi^2}\sum_i(-)^{F_i}M_i^4\ln
\frac{M_i^2}{\Lambda^2},
\label{Coleman}
\end{equation}
where the sum extends over all helicity states $i$,
$F_i$ and $M_i^2$ are the fermion number and mass
squared of the $i$th state and $\Lambda$ is a
renormalization mass scale. In order to use this
formula for creating a logarithmic slope which
drives the inflaton towards the minimum, one has
first to derive the mass spectrum of the model on
the `shifted' inflationary path.

\par
As mentioned, during inflation, $H^c$, $\bar{H}^c$
acquire constant values in the $\nu_H^c$,
$\bar{\nu}_H^c$ directions which are equal to
$v~(>0)$ and break $G_{PS}$ to $G_{S}$. We can
then write $\nu^c_H=v+\delta\nu^c_H$,
$\bar{\nu}^c_H=v+\delta\bar{\nu}^c_H$, where
$\delta\nu^c_H$, $\delta\bar{\nu}^c_H$ are complex
scalar fields. The (complex) deviations of the
fields $S$, $\phi$, $\bar{\phi}$ from their values
at a point on the `shifted' path (corresponding to
$S>0$) are similarly denoted as $\delta S$,
$\delta\phi$, $\delta\bar{\phi}$. We define the
complex scalar fields
\begin{eqnarray}
\theta &=&\frac{\delta\nu_H^c+\delta\bar{\nu}^c_H}
{\sqrt{2}},~~~~~\eta=\frac{\delta\nu_H^c-
\delta\bar{\nu}^c_H}{\sqrt{2}},\\
\zeta &=&\frac{\kappa\delta S+\lambda\delta\bar{\phi}
}{(\kappa^2+\lambda^2)^{\frac{1}{2}}},
~~~~~~\varepsilon=
\frac{\lambda\delta S-\kappa\delta\bar{\phi}}
{(\kappa^2+\lambda^2)^{\frac{1}{2}}}\cdot
\end{eqnarray}
We find that $\eta$ and $\varepsilon$ do not acquire
any masses from the scalar potential in
Eq.(\ref{potential}). Actually, $\varepsilon$ (and
its SUSY partner) remains massless even after
including the gauge interactions (see below). It
corresponds to the complex inflaton field $\Sigma=
(\lambda S-\kappa\bar{\phi})/(\kappa^2+\lambda^2)
^{1/2}$, which on the `shifted' path takes the form
$\Sigma=(\kappa^2+\lambda^2)^{1/2}S/\lambda$. So,
in this case, the real normalized inflaton field is
$\sigma=2^{1/2}(\kappa^2+\lambda^2)^{1/2}S/\lambda$.

\par
Contrary to $\eta$ and $\varepsilon$, the complex
scalars $\theta$, $\delta\phi$ and $\zeta$ acquire
masses from the potential in Eq.(\ref{potential}).
Expanding these scalars in real and imaginary parts
$\theta=(\theta_1+i\theta_2)/\sqrt{2}$, $\delta\phi
=(\delta\phi_1+i\delta\phi_2)/\sqrt{2}$, $\zeta
=(\zeta_1+i\zeta_2)/\sqrt{2}$, we find that
the mass squared matrices $M_+^2$ and $M_-^2$ of
$\theta_1$, $\delta\phi_1$, $\zeta_1$ and $\theta_2$,
$\delta\phi_2$, $\zeta_2$ are given by
\begin{equation}
M_{\pm}^2=M^2\left(\begin{array}{ccc}
                   a^2 &~~ ab &~~ 0 \\
                   ab &~~ b^2+c^2\pm f^2 &~~ -cb\\
                   0 &~~ -cb &~~ a^2 + b^2
                   \end{array} \right),
\label{M+-}
\end{equation}
where $a^2=2\kappa^2(1/4\xi+1)+\lambda^2/\xi$, $b^2=
\beta(\kappa^2+\lambda^2)/\kappa\xi$, $c^2=2\beta^2
\lambda^2\sigma^2/M^2(\kappa^2+\lambda^2)$, $f^2=2
\kappa\beta\lambda^2(1/4\xi-1)/(\kappa^2+\lambda^2)$
($a$, $b$, $c$, $f>0$).

\par
One can show that, for $\sigma\to\infty$
($c\to\infty$), all the eigenvalues of these two
mass squared matrices are positive. So, for large
values of $\sigma$, the `shifted' path is a valley
of local minima. As $\sigma$ decreases, one
eigenvalue may become negative destabilizing the
trajectory. From continuity, no eigenvalue can
become negative without passing from zero. So, the
critical (instability) point on the `shifted'
trajectory is encountered when one of the
determinants of the matrices in Eq.(\ref{M+-}),
which are $\det M_\pm^2=M^6a^2(a^2c^2\pm f^2(a^2+
b^2))$, vanishes. We see that $\det M_+^2$ is always
positive, while $\det M_-^2$ vanishes at $c^2=f^2(1+
b^2/a^2)$, which corresponds to the critical point
of the `shifted' path given by
\begin{equation}
\left(\frac{\sigma_c}{M}\right)^2=\frac{\kappa}{\beta}
\left(\frac{1}{4\xi}-1\right)\frac{2\kappa^2\left(1+
\frac{\kappa+2\beta}{4\kappa\xi}\right)+\frac{
\lambda^2(\kappa+\beta)}{\kappa\xi}}{2\kappa^2\left(1+
\frac{1}{4\xi}\right)+\frac{\lambda^2}{\xi}}\cdot
\label{sigmac}
\end{equation}

\par
The superpotential in Eq.(\ref{superpotential}) gives
rise to mass terms between the fermionic partners of
$\theta$, $\delta\phi$ and $\zeta$. The square of the
corresponding mass matrix is found to be
\begin{equation}
M_{0}^2=M^2\left(\begin{array}{ccc}
                   a^2 &~~ ab &~~ 0 \\
                   ab &~~ b^2+c^2 &~~ -cb\\
                   0 &~~ -cb &~~ a^2 + b^2
                   \end{array} \right).
\label{M0}
\end{equation}

\par
To complete the spectrum in the SM singlet sector,
which consists of the superfields $\nu^c_H$,
$\bar{\nu}^c_H$, $S$, $\phi$ and $\bar{\phi}$  (SM
singlet directions), we must consider the following
D-terms in the scalar potential:
\begin{equation}
\frac{1}{2}g^2\sum_{a}(H^c\,^{*}T^aH^c+
\bar{H}^c\,^{*}T^a\bar{H}^c)^2,
\label{Dterm}
\end{equation}
where $g$ is the $G_{PS}$ gauge coupling constant
and the sum extends over all the generators $T^a$
of $G_{PS}$. The part of this sum over the
generators $T^{15}=(1/2\sqrt{6})~\mathrm{diag}
(1,1,1,-3)$ of $SU(4)_c$ and $T^3=(1/2)~
\mathrm{diag}(1, -1)$ of $SU(2)_R$ gives rise to
a mass term for the normalized real scalar field
$\eta_1$ with $m^2=5g^2v^2/2$. The field $\eta_2$,
however, is left massless by the D-terms and is
absorbed by the gauge boson $A^{\perp}=-(3/5)^{1/2}
A^{15}+(2/5)^{1/2}A^{3}$ which becomes massive with
$m^2=5g^2v^2/2$ ($A^{15}$, $A^{3}$ are the gauge
bosons corresponding to $T^{15}$, $T^{3}$).

\par
Contributions to the fermion masses also arise from
the Lagrangian terms
\begin{equation}
i\sqrt{2}g\sum_{a}\lambda^a(H^c\,^{*}T^a \psi_{H^c}
+\bar{H}^c\,^{*}T^a\psi_{\bar{H}^c})+{\rm h.c.},
\label{fermion}
\end{equation}
where $\lambda^a$ is the gaugino corresponding to
$T^a$ and $\psi_{H^c}$, $\psi_{\bar{H}^c}$ represent
the chiral fermions in the superfields $H^c$,
$\bar{H}^c$. Concentrating again on $T^{15}$, $T^3$,
we obtain a Dirac mass term between the chiral
fermion in the $\eta$ direction and
$-i\lambda^{\perp}$ (with $\lambda^{\perp}$ being the
SUSY partner of $A^{\perp}$) with $m^2=5g^2v^2/2$.
Note that the SM singlet components of $\phi$ and
$\bar{\phi}$ do not contribute to bosonic and
fermionic couplings analogous to the ones in
Eqs.(\ref{Dterm}) and (\ref{fermion}) since they
commute with $T^{15}$ and $T^{3}$.

\par
This completes the analysis of the SM singlet sector
of the model. In summary, we found two groups of
three real scalars with mass squared matrices
$M_\pm^2$ and three two component fermions with mass
matrix squared $M_0^2$. Also, one Dirac fermion (with
four components), one gauge boson and one real scalar,
all of them having the same mass squared
$m^2=5g^2v^2/2$ and, thus, not contributing to the
one-loop radiative corrections. From
Eq.(\ref{Coleman}), we find that the contribution of
the SM singlet sector to the radiative corrections to
the potential along the `shifted' path is given by
\begin{equation}
\Delta V=\frac{1}{64\pi^2}{\rm Tr}\left(M_+^4
\ln\frac{M_+^2}{\Lambda^2}+M_-^4\ln\frac{M_-^2}
{\Lambda^2}-2M_0^4\ln\frac{M_0^2}{\Lambda^2}\right)
\cdot
\label{ssect}
\end{equation}
One can show that, in this sector, ${\rm Tr}M^2=0$
and ${\rm Tr}M^4=2M^4f^4$, which is
$\sigma$-independent and, thus, the generated slope
on the `shifted' path is $\Lambda$-independent.

\par
We now turn to the $u^c$ and $\bar{u}^c$ type
fields which are color antitriplets with charge
$-2/3$ and color triplets with charge $2/3$
respectively. Such fields exist in $H^c$,
$\bar{H}^c$, $\phi$ and $\bar\phi$, and we shall
denote them by $u^c_H$, $\bar{u}^c_H$, $u^c_{\phi}$,
$\bar{u}^c_{\phi}$, $u^c_{\bar\phi}$ and
$\bar{u}^c_{\bar\phi}$. The relevant expansion of
$\phi$ is
\begin{equation}
\phi=\left[\frac{1}{\sqrt{12}}\left(
\begin{array}{cccc}
1&~0&~0&~0\\
0&~1&~0&~0\\
0&~0&~1&~0\\
0&~0&~0&-3\end{array}\right),~\frac{1}{\sqrt{2}}
\left(\begin{array}{cc}
1&~0\\
0&-1\end{array}\right)\right]\phi+
\left(\begin{array}{cccc}
0&~0&~0&~0\\
0&~0&~0&~0\\
0&~0&~0&~0\\
1&~0&~0&~0\\
\end{array} \right)u^c_{\phi}
+\left(\begin{array}{cccc}
0&~0&~0&~1\\
0&~0&~0&~0\\
0&~0&~0&~0\\
0&~0&~0&~0\\
\end{array} \right)\bar{u}^c_{\phi}+\cdots,
\label{phiu}
\end{equation}
where the SM singlet in $\phi$ (denoted by the
same symbol) is also shown with the first
(second) matrix in the brackets belonging to
the algebra of $SU(4)_c$ ($SU(2)_R$). The fields
$u^c_{\phi}$, $\bar{u}^c_{\phi}$ are $SU(2)_R$
singlets, so only their $SU(4)_c$ structure is
shown and summation over their $SU(3)_c$ indices
is implied in the ellipsis. The field $\bar\phi$
can be similarly expanded.

\par
In the bosonic $u^c$, $\bar{u}^c$ type sector,
we find that the mass squared matrices
$m(u)_{\pm}^2$  of the complex scalars
$u_{H\pm}^c=(u_H^c\pm\bar{u}^{c~*}_H)/\sqrt{2}$,
$u_{\phi\pm}^c=(u_{\phi}^c\pm\bar{u}^{c~*}_{\phi})
/\sqrt{2}$ and $u_{\bar{\phi}\pm}^c=
(u_{\bar{\phi}}^c\pm\bar{u}_{\bar{\phi}}^{c~*})/
\sqrt{2}$ are given by
\begin{equation}
m(u)_{+}^2=M^2\left(\begin{array}{ccc}
\frac{4\kappa^2c^2}{9\beta^2}+
\frac{2\lambda^2a^2\beta}{3\kappa\xi b^2}+
\frac{2\kappa f^2}{3\beta}&~~
-\frac{\sqrt{2}\lambda^2\beta a}
{\sqrt{3}\kappa\xi b}
&~~-\frac{2\sqrt{2}\kappa^{\frac{1}{2}}
\lambda ca}{3\sqrt{3}\beta^{\frac{1}{2}}
\xi^{\frac{1}{2}}b}\\
-\frac{\sqrt{2}\lambda^2\beta a}{\sqrt{3}
\kappa\xi b}
&~~\frac{\beta\lambda^2}{\kappa\xi}+c^2-f^2&~~
-\frac{\lambda\beta^{\frac{1}{2}}c}
{\kappa^{\frac{1}{2}}\xi^{\frac{1}{2}}}\\
-\frac{2\sqrt{2}\kappa^{\frac{1}{2}}\lambda ca}
{3\sqrt{3}\beta^{\frac{1}{2}}\xi^{\frac{1}{2}}b}
&~~-\frac{\lambda\beta^{\frac{1}{2}}c}
{\kappa^{\frac{1}{2}}\xi^{\frac{1}{2}}}&
~~~\frac{\beta \lambda^2}{\kappa\xi}+
\frac{2\lambda^2a^2\beta}{3\kappa\xi b^2}
\end{array}\right)
\label{mu+}
\end{equation}
and
\begin{equation}
m(u)_{-}^2= M^2\left(\! \begin{array}{ccc}
\frac{4\kappa^2c^2}{9\beta^2} \!+\! \frac{2\lambda^2
a^2\beta}{3\kappa\xi b^2}\!-\! \frac{2\kappa f^2}
{3\beta}\!+\!\frac{g^2a^2\beta}{2\kappa\xi b^2} \!\!&
\frac{-\sqrt{2}\lambda^2\beta a}
{\sqrt{3}\kappa\xi b}\!+\! \frac{g^2a}{\sqrt{6}\xi b}
\!\!& \frac{-2\sqrt{2}\kappa^{\frac{1}{2}}\lambda ca}
{3\sqrt{3}\beta^{\frac{1}{2}}\xi^{\frac{1}{2}}b}
\!+\!\frac{g^2\kappa^{\frac{1}{2}}ac}{\sqrt{6}
\xi^{\frac{1}{2}}\beta^{\frac{1}{2}}\lambda b}\\
\!\frac{-\sqrt{2}\lambda^2\beta a}
{\sqrt{3}\kappa\xi b} \!+\! \frac{g^2a}
{\sqrt{6}\xi b}\!\!& \frac{\beta\lambda^2}
{\kappa\xi} \!+\! c^2 \!+\!
f^2\!+\! \frac{g^2\kappa}{3\beta\xi} \!\!&
\!\frac{-\lambda\beta^{\frac{1}{2}}c}
{\kappa^{\frac{1}{2}}\xi^{\frac{1}{2}}} \!+\!
\frac{g^2c\kappa^{\frac{3}{2}}}
{3\xi^{\frac{1}{2}}\beta^{\frac{3}{2}}\lambda}\\
\!\frac{-2\sqrt{2}\kappa^{\frac{1}{2}}\lambda ca}
{3\sqrt{3}\beta^{\frac{1}{2}}\xi^{\frac{1}{2}}b}
\!+\! \frac{g^2\kappa^{\frac{1}{2}}ac}{\sqrt{6}
\xi^{\frac{1}{2}} \beta^{\frac{1}{2}}\lambda b}
\!\!&\!\frac{-\lambda\beta^{\frac{1}{2}}c}
{\kappa^{\frac{1}{2}}\xi^{\frac{1}{2}}} \!+\!
\frac{g^2c\kappa^{\frac{3}{2}}}
{3\xi^{\frac{1}{2}}\beta^{\frac{3}{2}}\lambda}
\!\!&\frac{\beta\lambda^2}{\kappa\xi} \!+\!
\frac{2\lambda^2a^2\beta}{3\kappa\xi b^2} \!+\!
\frac{g^2\kappa^2c^2}{3\lambda^2\beta^2}
\end{array}\! \right).
\label{mu-}
\end{equation}
The mass squared matrix $m(u)^2_+$ has one zero
eigenvalue corresponding to the Goldstone boson
which is absorbed by the superhiggs mechanism.
This is easily checked by showing that
$\det m(u)_+^2=0$. However, it does no harm to
keep this Goldstone mode since it has vanishing
contribution to the radiative corrections in
Eq.(\ref{Coleman}) anyway.

\par
In the $u^c$, $\bar{u}^c$ type sector, we obtain
four Dirac fermions (per color) $\psi^D_{u^c}=
\psi_{u^c}+\psi_{\bar{u}^c}^c$,
$\psi^D_{u^c_{\phi}}=\psi_{u^c_{\phi}}+
\psi^c_{\bar{u}^c_{\phi}}$,
$\psi^D_{u^c_{\bar{\phi}}}
=\psi_{u^c_{\bar{\phi}}}+
\psi^c_{\bar{u}^c_{\bar{\phi}}}$, $-i\lambda^D=
-i(\lambda^{+}+\lambda^{-~c})$. Here,
$\lambda^\pm=(\lambda^1\pm i\lambda^2)/\sqrt{2}$,
where $\lambda^1$ ($\lambda^2$) is the gaugino
color triplet corresponding to the $SU(4)_c$
generators with $1/2$ ($-i/2$) in the $i4$ and
$1/2$ ($i/2$) in the $4i$ entry ($i=1,2,3$). The
fermionic mass matrix is
\begin{equation}
m(\psi_u)=M\left(\begin{array}{cccc}
\frac{2\kappa c}{3\beta}&~~0&~~-\frac{\sqrt{2}
\beta^{\frac{1}{2}}\lambda a}{\sqrt{3}
\kappa^{\frac{1}{2}}\xi^{\frac{1}{2}}b}&~~
\frac{ga\beta^{\frac{1}{2}}}{\sqrt{2}
\kappa^{\frac{1}{2}}\xi^{\frac{1}{2}}b}\\
0&~~-c&~~\frac{\beta^{\frac{1}{2}}\lambda}
{\kappa^{\frac{1}{2}}\xi^{\frac{1}{2}}}&~~
\frac{g \kappa^{\frac{1}{2}}}
{\sqrt{3}\beta^{\frac{1}{2}}\xi^{\frac{1}{2}}}\\
-\frac{\sqrt{2}\beta^{\frac{1}{2}}\lambda a}
{\sqrt{3}\kappa^{\frac{1}{2}}\xi^{\frac{1}{2}}b}
&~~\frac{\beta^{\frac{1}{2}}\lambda}
{\kappa^{\frac{1}{2}}\xi^{\frac{1}{2}}}&~~0&~~
\frac{g\kappa c}{\sqrt{3}\lambda\beta}\\
\frac{ga\beta^{\frac{1}{2}}}{\sqrt{2}
\kappa^{\frac{1}{2}}\xi^{\frac{1}{2}}b}&~~
\frac{g\kappa^{\frac{1}{2}}}{\sqrt{3}
\beta^{\frac{1}{2}}\xi^{\frac{1}{2}}}&~~
\frac{g\kappa c}{\sqrt{3}\lambda\beta}&~~ 0
\end{array} \right).
\label{mpsiu}
\end{equation}
To complete this sector, we must also include the
gauge bosons $A^\pm$ which are associated with
$\lambda^\pm$. They acquire a mass squared $m^2_g
=g^2M^2(a^2\beta/2\kappa\xi b^2+\kappa/3\beta\xi+
\kappa^2c^2/3\beta^2\lambda^2)$.

\par
The overall contribution of the $u^c$, $\bar{u}^c$
type sector to $\Delta V$ in Eq.(\ref{Coleman}) is
\begin{equation}
\Delta V=\frac{3}{32\pi^2}{\rm Tr}\left(m(u)^4_+
\ln\frac{m(u)_+^2}{\Lambda^2}+m(u)^4_-
\ln\frac{m(u)_-^2}{\Lambda^2}-2m(\psi_u)^4
\ln\frac{m(\psi_u)^2}{\Lambda^2}+3m_g^4
\ln\frac{m_g^2}{\Lambda^2}\right).
\label{usect}
\end{equation}
In this sector, ${\rm Tr}M^2=0$ and ${\rm Tr}M^4
=12M^4f^4(1+4\kappa^2/9\beta^2-2g^2\kappa^2/3
\beta^2\lambda^2)$. So, the contribution of this
sector to the slope of the `shifted' path is
also $\Lambda$-independent.

\par
We will now discuss the contribution from the $e^c$,
$\bar{e}^c$ type sector consisting of color singlets
with charge $1$, $-1$. Such fields exist in $H^c$,
$\bar{H}^c$, $\phi$, $\bar{\phi}$ and we denote
them by $e^c_H$, $\bar{e}^c_H$, $e^c_{\phi}$,
$\bar{e}^c_{\phi}$, $e^c_{\bar\phi}$,
$\bar{e}^c_{\bar\phi}$. The field $\phi$ can be
expanded in $e^c_{\phi}$, $\bar{e}^c_{\phi}$ as
follows:
\begin{equation}
\phi=\left[\frac{1}{\sqrt{12}}\left(
\begin{array}{cccc}
1&~0&~0&~0\\
0&~1&~0&~0\\
0&~0&~1&~0\\
0&~0&~0&-3\end{array}\right),~
\left(\begin{array}{cc}
0&~1\\
0&~0\end{array}\right)e^c_{\phi}+\left(
\begin{array}{cc}
0&~0\\
1&~0\end{array}\right)\bar{e}^c_{\phi}\right]+\cdots,
\label{phie}
\end{equation}
with the same notation as in Eq.(\ref{phiu}). A
similar expansion holds for $\bar\phi$. The analysis
in this sector is similar to the one in the $u^c$,
$\bar{u}^c$ type sector and we only summarize the
results.

\par
In the bosonic sector, we obtain two groups each
consisting of three complex scalars with mass
squared matrices
\begin{equation}
m(e)_{+}^2=M^2\left( \begin{array}{ccc}
\frac{\kappa^2c^2}{\beta^2}+
\frac{\lambda^2a^2\beta}
{\kappa\xi b^2}+\frac{\kappa f^2}{\beta}&~~
\frac{\lambda^2\beta a}{\kappa\xi b}&~~
\frac{\kappa^{\frac{1}{2}}\lambda ca}
{\beta^{\frac{1}{2}}\xi^{\frac{1}{2}}b}\\
\frac{\lambda^2\beta a}{\kappa\xi b}
&~~\frac{\beta\lambda^2}{\kappa\xi}+c^2-f^2&~~
-\frac{\lambda \beta^{\frac{1}{2}}c}
{\kappa^{\frac{1}{2}}\xi^{\frac{1}{2}}}\\
\frac{\kappa^{\frac{1}{2}}\lambda ca}
{\beta^{\frac{1}{2}}\xi^{\frac{1}{2}}b}
&~~-\frac{\lambda\beta^{\frac{1}{2}}c}
{\kappa^{\frac{1}{2}}\xi^{\frac{1}{2}}}&~~
\frac{\beta\lambda^2}{\kappa\xi}+
\frac{\lambda^2a^2\beta}{\kappa\xi b^2}
\end{array} \right)
\label{me+}
\end{equation}
and
\begin{equation}
m(e)_{-}^2=M^2\left(\! \begin{array}{ccc}
\frac{\kappa^2c^2}{\beta^2} \!+\!
\frac{\lambda^2a^2\beta}{\kappa\xi b^2}\!-\!
\frac{\kappa f^2}{\beta}\!+\!\frac{g^2a^2\beta}
{2\kappa\xi b^2} \!\!&\frac{\lambda^2\beta a}
{\kappa\xi b}\!-\! \frac{g^2a}{2\xi b} \!\!&
\frac{\kappa^{\frac{1}{2}}\lambda ca}
{\beta^{\frac{1}{2}}\xi^{\frac{1}{2}}b} \!-\!
\frac{g^2\kappa^{\frac{1}{2}}ac}{2\xi^{\frac{1}{2}}
\beta^{\frac{1}{2}}\lambda b}\\
\!\frac{\lambda^2\beta a}{\kappa\xi b} \! - \!
\frac{g^2a}{2\xi b} \!\!& \frac{\beta\lambda^2}
{\kappa\xi} \!+\! c^2 \!+\!f^2\!+\! \frac{g^2\kappa}
{2\xi\beta} \!\!&\!\frac{-\lambda\beta^{\frac{1}{2}}c}
{\kappa^{\frac{1}{2}}\xi^{\frac{1}{2}}} \!+\!
\frac{g^2\kappa^{\frac{3}{2}}c}{2\xi^{\frac{1}{2}}
\beta^{\frac{3}{2}}\lambda}\\
\!\frac{\kappa^{\frac{1}{2}}\lambda ca}
{\beta^{\frac{1}{2}} \xi^{\frac{1}{2}} b}\!-\!
\frac{g^2\kappa^{\frac{1}{2}} ac}{2\xi^{\frac{1}{2}}
\beta^{\frac{1}{2}}\lambda b}
\!\!&\!\frac{-\lambda\beta^{\frac{1}{2}}c}
{\kappa^{\frac{1}{2}}\xi^{\frac{1}{2}}} \!+\!
\frac{g^2\kappa^{\frac{3}{2}}c}{2\xi^{\frac{1}{2}}
\beta^{\frac{3}{2}}\lambda} \!\!&
\frac{\beta\lambda^2}{\kappa\xi} \!+\!
\frac{\lambda^2a^2\beta}{\kappa\xi b^2}
\!+\!\frac{g^2\kappa^2c^2}{2\lambda^2\beta^2}
\end{array}\! \right).
\label{me-}
\end{equation}
The matrix $m(e)^2_+$, similarly to $m(u)^2_+$ in
the $u^c$, $\bar{u}^c$ type sector, has one zero
eigenvalue corresponding to the Goldstone mode
absorbed by the superhiggs mechanism.

\par
In the fermion sector, we obtain four Dirac fermions
with mass matrix given by
\begin{equation}
m(\psi_e)=M\left(\begin{array}{cccc}
\frac{\kappa c}{\beta}&~~0 &~~
\frac{\beta^{\frac{1}{2}}\lambda a}
{\kappa^{\frac{1}{2}}\xi^{\frac{1}{2}}b}&~~
\frac{ga\beta^{\frac{1}{2}}}{\sqrt{2}
\kappa^{\frac{1}{2}}\xi^{\frac{1}{2}}b}\\
0&~~-c&~~\frac{\beta^{\frac{1}{2}}\lambda}
{\kappa^{\frac{1}{2}}\xi^{\frac{1}{2}}}&~~
-\frac{g\kappa^{\frac{1}{2}}}
{\sqrt{2}\beta^{\frac{1}{2}}\xi^{\frac{1}{2}}}\\
\frac{\beta^{\frac{1}{2}}\lambda a}
{\kappa^{\frac{1}{2}}\xi^{\frac{1}{2}}b}
&~~\frac{\beta^{\frac{1}{2}}\lambda}
{\kappa^{\frac{1}{2}}\xi^{\frac{1}{2}}}&~~0
&~~-\frac{g\kappa c}{\sqrt{2}\lambda\beta}\\
\frac{ga\beta^{\frac{1}{2}}}{\sqrt{2}
\kappa^{\frac{1}{2}}\xi^{\frac{1}{2}}b}&~~
-\frac{g\kappa^{\frac{1}{2}}}
{\sqrt{2}\beta^{\frac{1}{2}}\xi^{\frac{1}{2}}}
&~~-\frac{g\kappa c}{\sqrt{2}\lambda\beta}&~~0
\end{array}\right).
\label{mpsie}
\end{equation}
We also have, in this sector, one complex gauge
boson with mass squared given by $\hat{m}_{g}^2=
g^2M^2(a^2\beta/2\kappa\xi b^2+\kappa/2\beta\xi
+\kappa^2c^2/2\beta^2\lambda^2)$.

\par
The contribution of the $e^c$, $\bar{e}^c$ type
sector to $\Delta V$ is
\begin{equation}
\Delta V=\frac{1}{32\pi^2}{\rm Tr}\left(m(e)^4_+
\ln\frac{m(e)_+^2}{\Lambda^2}+m(e)^4_-
\ln\frac{m(e)_-^2}{\Lambda^2}- 2m(\psi_e)^4
\ln\frac{m(\psi_e)^2}{\Lambda^2}+3\hat{m}_{g}^4
\ln \frac{\hat{m}_{g}^2}{\Lambda^2}\right).
\label{esect}
\end{equation}
One can show that ${\rm Tr}M^2=0$ and ${\rm Tr}
M^4=4M^4f^4(1+\kappa^2/\beta^2-g^2\kappa^2/
\beta^2\lambda^2)$ in this sector and, thus,
its contribution to the inflationary slope is
again $\Lambda$-independent.

\par
We next consider the $d^c$ and $\bar{d}^c$ type
sector consisting of color antitriplets with
charge $1/3$ and color triplets with charge
$-1/3$. We have the fields $d^c_H$, $\bar{d}^c_H$
from $H^c$, $\bar{H}^c$ and the fields $d^c_{\phi}$,
$\bar{d}^c_{\phi}$, $d^c_{\bar\phi}$,
$\bar{d}^c_{\bar\phi}$ from $\phi$, $\bar{\phi}$.
Note that $\phi$ can be expanded as
\begin{equation}
\phi=\left[\left(\begin{array}{cccc}
0&~0&~0&~0\\
0&~0&~0&~0\\
0&~0&~0&~0\\
1&~0&~0&~0\end{array}\right),
\left(\begin{array}{cc}
0&~1\\
0&~0\end{array}\right)\right]d^c_{\phi}
+\left[\left(\begin{array}{cccc}
0&~0&~0&~1\\
0&~0&~0&~0\\
0&~0&~0&~0\\
0&~0&~0&~0\end{array}\right),
\left(\begin{array}{cc}
0&~0\\
1&~0\end{array}\right)\right]\bar{d}^c_{\phi}+\cdots,
\label{phid}
\end{equation}
with the notation of Eq.(\ref{phiu}). The field
$\bar\phi$ is similarly expanded. In order to give
superheavy masses to $d^c_H$ and $\bar{d}^c_H$, we
introduce \cite{leontaris} a $SU(4)_c$ 6-plet
superfield $G=(6,1,1)$ with the superpotential
couplings $xGH^cH^c$, $yG\bar{H}^c\bar{H}^c$. The
field $G$ splits, under $G_S$, into
$g^c=(\bar{3},1,1/3)$ and $\bar{g}^c=(3,1,-1/3)$.

\par
The mass terms of the complex scalars $d^c_H$,
$\bar{d}^c_H$, $d^c_{\phi}$, $\bar{d}^c_{\phi}$,
$d^c_{\bar\phi}$, $\bar{d}^c_{\bar\phi}$, $g^c$,
$\bar{g}^c$ are
\begin{eqnarray}
\mathcal{L}_m(d)&=&M^2\Big[\Big(\frac{\kappa^2c^2}
{9\beta^2}+\frac{2a^2\beta}{\kappa\xi b^2}
\Big(\frac{2\lambda^2}{3}+x^2\Big)\Big)
\vert d^c_H\vert^2+\Big(\frac{\kappa^2c^2}
{9\beta^2}+\frac{2a^2\beta}{\kappa\xi b^2}
\Big(\frac{2\lambda^2}{3}+y^2\Big)\Big)
\vert\bar{d}^c_H\vert^2
\nonumber \\
&+&\Big(\frac{\beta\lambda^2}{\kappa\xi}+c^2\Big)
(\vert d^c_{\phi}\vert^2+
\vert\bar{d}^c_{\phi}\vert^2)+
\Big(\frac{\beta\lambda^2}{\kappa\xi}+
\frac{4\lambda^2a^2\beta}{3\kappa\xi b^2}\Big)
(\vert d^c_{\bar\phi}\vert^2+
\vert\bar{d}^c_{\bar\phi}\vert^2)
\nonumber\\
&+&\frac{2a^2\beta^2}{\kappa\xi b^2}
(y^2\vert g^c\vert^2+x^2\vert\bar{g}^c\vert^2)
+\Big(\frac{\kappa f^2}{3\beta}d^c_H\bar{d}^c_H
-\frac{2\lambda^2\beta a}
{\sqrt{3}\kappa\xi b}(d^c_Hd^{c~*}_{\phi}+
\bar{d}^c_H\bar{d}^{c~*}_{\phi})
\nonumber\\
&-&\frac{2\kappa^{\frac{1}{2}}\lambda c a}
{3\sqrt{3}\beta^{\frac{1}{2}}\xi^{\frac{1}{2}} b}
(d^c_Hd^{c~*}_{\bar{\phi}}+
\bar{d}^c_H\bar{d}^{c~*}_{\bar{\phi}})
-\frac{\sqrt{2}\kappa^{\frac{1}{2}} c a}
{3\beta^{\frac{1}{2}}\xi^{\frac{1}{2}} b}
(yd^c_Hg^c\,^{*}+x\bar{d}^c_H\bar{g}^c\,^{*})
-f^2d^c_{\phi}\bar{d}^c_{\phi}
\nonumber\\
&-&\frac{\lambda\beta^{\frac{1}{2}}c}
{\kappa^{\frac{1}{2}}\xi^{\frac{1}{2}}}
(d^c_{\phi} d^{c~*}_{\bar{\phi}}+
\bar{d}^c_{\phi} \bar{d}^{c~*}_{\bar{\phi}})
+\frac{2\sqrt{2}\lambda a^2\beta}
{\sqrt{3}\kappa\xi b^2}
(y d^c_{\bar{\phi}}g^c\,^{*}+
x\bar{d}^c_{\bar{\phi}}
\bar{g}^c\,^{*})+{\rm h.c.}\Big)\Big].
\end{eqnarray}
From these mass terms, we can construct the
$8\times 8$ mass squared matrix $m(d)^2$ of
the complex scalar fields $d^c_H$,
$\bar{d}^{c~*}_H$, $d^c_{\phi}$,
$\bar{d}^{c~*}_{\phi}$, $d^c_{\bar{\phi}}$,
$\bar{d}^{c~*}_{\bar{\phi}}$,
$g^c$, $\bar{g}^c\,^{*}$.

\par
In the fermion sector, we obtain four Dirac
fermions per color with mass matrix
\begin{equation}
m(\psi_d)=M\left(\begin{array}{cccc}
\frac{\kappa c}{3\beta}&~~0&~~
-\frac{2\beta^{\frac{1}{2}}\lambda a}
{\sqrt{3}\kappa^{\frac{1}{2}}\xi^{\frac{1}{2}}b}
&~~-\frac{\sqrt{2}\beta^{\frac{1}{2}}ax}
{\kappa^{\frac{1}{2}}\xi^{\frac{1}{2}}b}\\
0&~~-c&~~\frac{\beta^{\frac{1}{2}}\lambda}
{\kappa^{\frac{1}{2}}\xi^{\frac{1}{2}}}&~~0\\
-\frac{2\beta^{\frac{1}{2}}\lambda a}
{\sqrt{3}\kappa^{\frac{1}{2}}\xi^{\frac{1}{2}} b}
&~~\frac{\beta^{\frac{1}{2}}\lambda}
{\kappa^{\frac{1}{2}}\xi^{\frac{1}{2}}}&~~0&~~0\\
-\frac{\sqrt{2}\beta^{\frac{1}{2}}ay}
{\kappa^{\frac{1}{2}}\xi^{\frac{1}{2}}b}&~~0&~~0
&~~0\end{array}\right).
\label{mpsid}
\end{equation}
Note that there are no D-terms, gauge bosons or
gauginos in this sector.

\par
The contribution of the $d^c$, $\bar{d}^c$ type
sector to $\Delta V$ is given by
\begin{equation}
\Delta V=\frac{3}{32\pi^2}{\rm Tr}\left(m(d)^4
\ln\frac{m(d)^2}{\Lambda^2}-
2(m(\psi_d)m(\psi_d)^{\dagger})^2
\ln\frac{m(\psi_d)m(\psi_d)^{\dagger}}{\Lambda^2}
\right).
\label{dsect}
\end{equation}
We find that ${\rm Tr} M^2=0$ and ${\rm Tr}M^4=12
M^4f^4(1+\kappa^2/9\beta^2)$, in this sector,
implying that its contribution to the inflationary
slope is again $\Lambda$-independent.

\par
Finally, we consider the $q^c$ and $\bar{q}^c$ type
superfields which are color antitriplets with
charge $-5/3$ and color triplets with charge $5/3$.
They exist in $\phi$, $\bar{\phi}$ and we call them
$q^c_{\phi}$, $\bar{q}^c_{\phi}$, $q^c_{\bar{\phi}}$,
$\bar{q}^c_{\bar{\phi}}$. The relevant expansion of
$\phi$ is
\begin{equation}
\phi=\left[\left(\begin{array}{cccc}
0&~0&~0&~0\\
0&~0&~0&~0\\
0&~0&~0&~0\\
1&~0&~0&~0\end{array}\right),
\left(\begin{array}{cc}
0&~0\\
1&~0 \end{array}\right)\right]q^c_{\phi}+
\left[\left(\begin{array}{cccc}
0&~0&~0&~1\\
0&~0&~0&~0\\
0&~0&~0&~0\\
0&~0&~0&~0\end{array} \right),
\left(\begin{array}{cc}
0&~1\\
0&~0\end{array}\right)\right]\bar{q}^c_{\phi}
+\cdots,
\label{phiq}
\end{equation}
with the notation of Eq.(\ref{phiu}). A similar
expansion holds for $\bar\phi$.

\par
We find that the mass squared matrices in the $q^c$,
$\bar{q}^c$ type bosonic sector are given by
\begin{equation}
m(q)_{\pm}^2=M^2\left(\begin{array}{cc}
\frac{\beta\lambda^2}{\kappa\xi}+c^2\mp f^2&~~
-\frac{\lambda\beta^{\frac{1}{2}}c}
{\kappa^{\frac{1}{2}}\xi^{\frac{1}{2}}}\\
-\frac{\lambda \beta^{\frac{1}{2}}c}
{\kappa^{\frac{1}{2}}\xi^{\frac{1}{2}}}&~~
\frac{\beta\lambda^2}{\kappa\xi}
\end{array}\right).
\label{mq+-}
\end{equation}

\par
The fermion mass matrix in this sector is given by
\begin{equation}
m(\psi_q)=M\left(\begin{array}{cc}
-c&~~\frac{\beta^{\frac{1}{2}}\lambda}
{\kappa^{\frac{1}{2}}\xi^{\frac{1}{2}}}\\
\frac{\beta^{\frac{1}{2}}\lambda}
{\kappa^{\frac{1}{2}}\xi^{\frac{1}{2}}}&~~ 0
\end{array}\right).
\label{mpsiq}
\end{equation}

\par
Furthermore, in $\phi$, $\bar{\phi}$, there exist
color octet, $SU(2)_R$ triplet superfields:
$\phi^0_8$, $\phi^\pm_8$, $\bar{\phi}^0_8$,
$\bar{\phi}^\pm_8$ with charge $0$, $1$, $-1$ as
indicated. It turns out that the mass (squared)
matrices in this sector are the same as the ones
in the $q^c$, $\bar{q}^c$ sector (see
Eqs.(\ref{mq+-}) and (\ref{mpsiq})).

\par
The combined contribution from the $q^c$,
$\bar{q}^c$ type and color octet fields to
$\Delta V$ is
\begin{equation}
\Delta V=\frac{15}{32\pi^2}{\rm Tr}\left(m(q)_+^4
\ln\frac{m(q)_+^2}{\Lambda^2}+
m(q)_-^4\ln\frac{m(q)_-^2}{\Lambda^2}-
2m(\psi_q)^4\ln\frac{m(\psi_q)^2}{\Lambda^2}
\right).
\label{qphisect}
\end{equation}
Of course, ${\rm Tr}M^2$ is vanishing in this
combined sector too and ${\rm Tr}M^4=60M^4f^4$,
so that we again have a $\Lambda$-independent
contribution to the inflationary slope.

\par
The final overall $\Delta V$ is found by adding
the contributions from the SM singlet sector in
Eq.(\ref{ssect}), the $u^c$, $\bar{u}^c$ type
sector in Eq.(\ref{usect}), the $e^c$,
$\bar{e}^c$ type sector in Eq.(\ref{esect}),
the $d^c$, $\bar{d}^c$ type sector in
Eq.(\ref{dsect}), and the combined $q^c$,
$\bar{q}^c$ type and color octet sector in
Eq.(\ref{qphisect}). These one-loop
radiative corrections are added to $V_0$
yielding the effective potential $V(\sigma)$
along the `shifted' inflationary trajectory.
They generate a slope on this trajectory which
is necessary for driving the system towards
the vacuum. The overall ${\rm Tr}M^4=2M^4f^4
(45+16\kappa^2/3\beta^2-6g^2\kappa^2/\beta^2
\lambda^2)$. This implies that the overall
slope is $\Lambda$-independent. This is, in
fact, a crucial property of the model since
otherwise observable quantities like the
quadrupole anisotropy $(\delta T/T)_Q$ of the
CMBR or the spectral index would depend on
the scale $\Lambda$ which remains undetermined.

\par
The slow roll parameters are given by
(see e.g., Ref.\cite{lectures})
\begin{equation}
\epsilon\simeq\frac{m_P^2}{2}~\left(
\frac{V'(\sigma)}{V_0}\right)^2,
~~~~~~~~\eta\simeq m_P^2~\frac{V''(\sigma)}
{V_0},
\label{slowroll}
\end{equation}
where the primes denote derivation with respect
to the real normalized inflaton field $\sigma$
and $m_P\simeq 2.44\times 10^{18}~{\rm GeV}$ is
the `reduced' Planck scale. The conditions for
inflation to take place are $\epsilon\leq 1$ and
$\vert\eta\vert\leq 1$.

\par
The number of e-foldings  our present horizon
scale suffered during inflation can be calculated
as follows (see e.g., Ref.\cite{lectures}):
\begin{equation}
N_Q\simeq\frac{1}{m_P^2}\int_{\sigma_f}
^{\sigma_Q}\frac{V_0}{V'(\sigma)}d\sigma,
\label{NQ}
\end{equation}
where $\sigma_f$ is the value of $\sigma$ at
the end of inflation and $\sigma_Q$ the value of
$\sigma$ when our present horizon scale crossed
outside the inflationary horizon. From
Ref.\cite{lectures}, one deduces that $N_Q$
should coincide with
\begin{equation}
N_Q\simeq\ln\left[4.41\times 10^{11}~
T_r^{\frac{1}{3}}~V_0^{\frac{1}{6}}
\right],
\label{efold}
\end{equation}
where the `reheat' temperature $T_r$ and the
inflationary scale $V_0^{1/4}$ are measured in
GeV. We will take $T_r\simeq 10^9~{\rm GeV}$,
which saturates the gravitino constraint
\cite{gravitino}.

\par
As can be easily seen from the relevant
expressions above, the effective potential
$V(\sigma)$ depends on the following parameters:
$M$, $m$, $\kappa$, $\beta$, $\lambda$ and $g$.
We fix the gauge coupling constant at $M_{GUT}$
to the value $g=0.7$, which leads to the correct
values of the SM gauge coupling constants at
$M_Z$. We also assume that the VEV $v_0=\langle
H^c\rangle=\langle \bar{H}^c\rangle $ at the
SUSY vacuum is equal to the SUSY GUT scale $M_{GUT}
\simeq 2.86\times 10^{16}~{\rm GeV}$. This allows
us to determine the mass scale $M$ in terms of the
parameters $m$, $\kappa$, $\beta$ and $\lambda$.
However, we find that the requirement that $M$ is
real restricts the possible values of these
parameters. For instance, $\lambda\lesssim 5\times
10^{-3}$ for $m\simeq 10^{16}~{\rm GeV}$, $\kappa
\simeq 10^{-3}$ and $\beta\simeq 1$.

\FIGURE{\epsfig{figure=sgc.eps,height=7.5cm}
\medskip
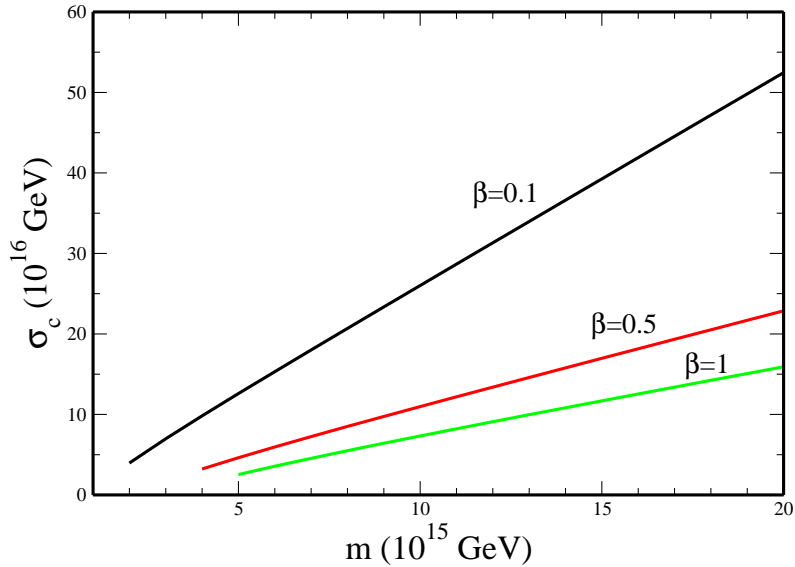
\caption{The critical value $\sigma_c$ of the
inflaton field as a function of the mass
parameter $m$ for $\kappa=\lambda=3\times
10^{-3}$ and $\beta=0.1$, $0.5$ and $1$, as
indicated on the curves.
\label{sgc}}}

\par
In figure \ref{sgc}, we present the critical
value $\sigma_c$ of the inflaton field, defined
in Eq.(\ref{sigmac}), as a function of the mass
scale $m$ for $\kappa=\lambda=3\times 10^{-3}$
and $\beta=0.1$, $0.5$ and $1$. As can be seen
from this figure, the smallest values of
$\sigma_c$ correspond to $\beta=1$. However,
in this case, the mass scale $m$ has to be
$\gtrsim 5\times 10^{15}~{\rm GeV}$ to avoid
complex values of $M$. The value of the
inflaton field $\sigma_f$ at which inflation
terminates cannot be smaller than its critical
value $\sigma_c$ where the `shifted' path
becomes unstable anyway. Thus, in order to
reduce the effect of SUGRA corrections which
could spoil \cite{cop,stew} the flatness of the
inflationary path, one would be tempted to
choose values for the parameters which minimize
$\sigma_c$. A possible set of such values is
$m=5\times 10^{15}~{\rm GeV}$, $\kappa=\lambda
=3\times 10^{-3}$ and $\beta=1$, which yield
$\sigma_c\simeq 4\times 10^{16}~{\rm GeV}$.
However, in this case, the condition
$\vert\eta\vert=1$ implies that inflation ends
at $\sigma_f\simeq 1.5\times 10^{18}~{\rm GeV}$,
which is quite large. Moreover, we find that
$\sigma_Q\simeq 1.6\times 10^{19}~{\rm GeV}$,
which is much bigger than $m_P$ and, thus,
this case is unacceptable.

\par
A better set of values is $m=2\times 10^{15}~
{\rm GeV}$, $\kappa=\lambda=5\times 10^{-3}$
and $\beta=0.1$, which also yield $\sigma_c
\simeq 4\times 10^{16}~{\rm GeV}$. In this
case, $\sigma_f\simeq 1.7\times
10^{17}~{\rm GeV}$ and $\sigma_Q\simeq 1.6
\times 10^{18}~{\rm GeV}$, which are much
smaller but still close to $m_P$. Values of
$\beta$ smaller than $0.1$ (with suitable
values of the other parameters) give also
very similar results. Actually, we find that
a general feature of our new shifted hybrid
inflationary model is that the relevant part
of inflation occurs at large values of
$\sigma$ which are close to $m_P$. (Note
that this property does not depend on the
value of $(\delta T/T)_Q$.) Consequently,
we are obliged to consider the SUGRA
corrections to the scalar potential and
invoke some mechanism to ensure that the
`shifted' inflationary path remains flat. We
will address this issue in the next section.

\par
We will now turn to the discussion of the
constraints imposed on the parameter space by
the measurements of the cosmic background
explorer (COBE) on the quadrupole anisotropy
$(\delta T/T)_Q$ of the CMBR . The $95\%$
confidence level allowed range of
$(\delta T/T)_Q$, under the condition that the
spectral index $n=1$, is given by \cite{cobe}
\begin{equation}
5.4\times 10^{-6}\lesssim \left(\frac{\delta T}
{T}\right)_Q\lesssim 7.8\times 10^{-6}.
\label{95cl}
\end{equation}
The quadrupole anisotropy can be calculated as
follows (see e.g., Ref.\cite{lectures}):
\begin{equation}
\left(\frac{\delta T}{T}\right)_Q\simeq\frac{1}
{12\sqrt{5}}\frac{V_0^{\frac{3}{2}}}
{V'(\sigma_Q)m_P^3}\cdot
\label{anisotropy}
\end{equation}
For a fixed $(\delta T/T)_Q$ in the
range in Eq.(\ref{95cl}), we can determine one
of the free parameters (say $\beta$) in terms
of the others ($m$, $\kappa$ and $\lambda$).
For instance, $(\delta T/T)_Q\simeq 6.6\times
10^{-6}$ corresponds to $\beta=0.1$ if $m=4.35
\times 10^{15}~{\rm GeV}$ and $\kappa=\lambda=
3\times 10^{-2}$. In this case, $\sigma_c\simeq
3.55\times 10^{16}~{\rm GeV}$, $\sigma_f\simeq
1.7\times 10^{17}~{\rm GeV}$ and $\sigma_Q
\simeq 1.6\times 10^{18}~{\rm GeV}$. Also,
$M\simeq 2.66\times 10^{16}~{\rm GeV}$, $N_Q
\simeq 57.7$ and $n\simeq 0.98$. We see that
the COBE constraint can be easily satisfied
with natural values of the parameters.
Moreover, superheavy SM non-singlets with
masses $\ll M_{GUT}$, which could disturb the
unification of the SM gauge couplings, are not
encountered.

\section{Supergravity corrections}
\label{sec:sugra}

\par
As we emphasized, the new shifted hybrid inflation
occurs at values of $\sigma$ which are quite close
to the `reduced' Planck scale. Thus, one cannot
ignore the SUGRA corrections to the scalar
potential. The scalar potential in SUGRA, without
the D-terms, is given by
\begin{equation}
V=e^{K/m_P^2}\left[(F_i)^*K^{i^*j}
F_j-3\frac{\vert W\vert^2}{m_P^2}\right],
\label{sugra}
\end{equation}
where $K$ is the K\"{a}hler potential, $F_i=W_i
+K_iW/m_P^2$, a subscript $i$ ($i^*$) denotes
derivation with respect to the complex scalar
field $\phi^i$ ($\phi^i\,^{*}$) and $K^{i^*j}$
is the inverse of the matrix $K_{ji^*}$.

\par
Consider a (complex) inflaton $\Sigma$
corresponding to a flat direction of global
SUSY with $W_{i\Sigma}=0$. We assume that the
potential on this path depends only on
$|\Sigma|$, which holds in our model due to a global
symmetry. From Eq.(\ref{sugra}), we find that
the SUGRA corrections lift the flatness of the
$\Sigma$ direction by generating a mass squared
for $\Sigma$ (see e.g., Ref.\cite{lyth})
\begin{equation}
m_\Sigma^2=\frac{V_0}{m_P^2}-
\frac{\vert W_\Sigma\vert^2}{m_P^2}+\sum_{i,j}
(W_i)^*K^{i^*j}_{\Sigma^*\Sigma}W_j+\cdots,
\label{mSigma}
\end{equation}
where the right hand side (RHS) is evaluated
on the flat direction with the explicitly
displayed terms taken at $\Sigma=0$. The ellipsis
represents higher order terms which are
suppressed by powers of $|\Sigma|/m_P$. The
slow roll parameter $\eta$ then becomes
\begin{equation}
\eta=1-\frac{\vert W_\Sigma\vert^2}{V_0}+
\frac{m_P^2}{V_0}\sum_{i,j}(W_i)^*
K^{i^*j}_{\Sigma^*\Sigma}W_j+\cdots,
\label{etasugra}
\end{equation}
which, in general, could be of order unity and,
thus, invalidate \cite{cop,stew} inflation. This
is the well-known $\eta$ problem of inflation in
local SUSY. Several proposals have been made
in the literature to overcome this difficulty
(for a review see e.g., Ref.\cite{lyth}).

\par
In standard and shifted SUSY hybrid inflation,
there is an automatic mutual cancellation
between the first two terms in the RHS of
Eq.(\ref{etasugra}). This is due to the fact
that $W_n=0$ on the inflationary path for all
field directions $n$ which are perpendicular to
this path, which implies that $|W_\Sigma|^2=V_0$
on the path. This is an important feature of
these models since, in general, the sum of the
first two terms in the RHS of Eq.(\ref{etasugra})
is positive and of order unity, thereby ruining
inflation. It is easily checked that these
properties persist in our new shifted hybrid
inflationary model too. In particular, the
superpotential on the `shifted' inflationary
path of this model takes the form
$W=V_0^{1/2}\Sigma$.

\par
In all these hybrid inflationary models, the
only non-zero contribution from the sum which
appears in the RHS of Eq.(\ref{etasugra})
originates from the term with $i=j=\Sigma$
(recall that $W_n=0$ on the path). This
contribution is equal to the dimensionless
coefficient of the quartic term
$|\Sigma|^4/4m_P^2$ in the K\"{a}hler potential
$K$. For inflation to remain intact, we need to
assume \cite{inf} that this coefficient is
somewhat small. The remaining terms give
negligible contributions to $\eta$ provided
$|\Sigma|\ll m_P$. The latter is true for
standard and shifted hybrid inflation. So, we
see that, in these models, a mild tuning of
just one parameter is \cite{inf} adequate for
protecting inflation from SUGRA corrections.

\par
In our present model, however, inflation takes
place at values of $|\Sigma|$ close to $m_P$.
So, the terms in the ellipsis in the RHS of
Eq.(\ref{etasugra}) cannot be ignored and may
easily invalidate inflation. We, thus, need to
invoke here a mechanism which can ensure that
the SUGRA corrections do not lift the flatness
of the inflationary path to all orders. A
suitable scheme has been suggested in
Ref.\cite{panag}. It has been argued that
special forms of the K\"{a}hler potential can
lead to the cancellation of the SUGRA
corrections which spoil slow roll inflation to
all orders. In particular, a specific form of
$K(\Sigma)$ (used in no-scale SUGRA models) was
employed and a gauge singlet field $Z$ with a
similar $K(Z)$ was introduced. It was pointed
out that, by assuming a superheavy VEV for the
$Z$ field through D-terms, an exact
cancellation of the inflaton mass on the
inflationary trajectory can be achieved.

\par
The mechanism of Ref.\cite{panag} can be
readily incorporated in our model to ensure
that the SUGRA corrections do not lift the
flatness of the inflationary path. The only
alteration caused to the Lagrangian along
this path is that the kinetic term of
$\sigma$ is now non-minimal. This affects
the equation of motion of $\sigma$ and,
consequently, the slow roll conditions,
$(\delta T/T)_Q$ and $N_Q$. The form of the
K\"{a}hler potential for $\Sigma$ used in
Ref.\cite{panag} is given by
\begin{equation}
K(\vert\Sigma\vert^2)=-Nm_P^2\ln\left(1-
\frac{\vert\Sigma\vert^2}{Nm_P^2}\right),
\label{kaehler}
\end{equation}
where $N=1$ or $2$. Here we take $N=2$. In
this case, the kinetic term of the real
normalized inflaton field $\sigma$ (recall
that $|\Sigma|=\sigma/\sqrt{2}$) is $(1/2)
(\partial^2 K/\partial\Sigma\partial\Sigma^*)
\dot{\sigma}^2$, where the overdot denotes
derivation with respect to the cosmic time
$t$ and $\partial^2K/\partial\Sigma\partial
\Sigma^*=(1-\sigma^2/2Nm_P^2)^{-2}$. Thus,
the Lagrangian on the `shifted' path is
given by
\begin{equation}
L=\int_{-\infty}^{\infty}dt\int d^3x~a^3(t)
\left[\frac{1}{2}\dot{\sigma}^2\left(1-
\frac{\sigma^2}{2 N m_P^2}\right)^{-2}-
V(\sigma)\right],
\label{lagrangian}
\end{equation}
where $a(t)$ is the scale factor of the universe.

\par
The evolution equation of $\sigma$ is found by
varying this Lagrangian with respect to $\sigma$:
\begin{equation}
\left[\ddot{\sigma}+3H\dot{\sigma}+\dot{\sigma}^2
\left(1-\frac{\sigma^2}{2Nm_P^2}\right)^{-1}
\frac{\sigma}{Nm_P^2}\right]\left(1-\frac{\sigma^2}
{2Nm_P^2}\right)^{-2}+V'(\sigma)=0,
\label{motion}
\end{equation}
where $H$ is the Hubble parameter. During
inflation, the `friction' term $3 H\dot{\sigma}$
dominates over the other two terms in the brackets
in Eq.(\ref{motion}). Thus, this equation reduces
to the `modified' inflationary equation
\begin{equation}
\dot{\sigma}=-\frac{V'(\sigma)}{3H}\left(1-
\frac{\sigma^2}{2 N m_P^2}\right)^2.
\label{infeq}
\end{equation}
Note that, for $\sigma\ll\sqrt{2N}m_P$, this
equation reduces to the standard inflationary
equation.

\par
To derive the slow roll conditions, we evaluate
the sum of the first and the third term in the
brackets in Eq.(\ref{motion}) by using
Eq.(\ref{infeq}):
\begin{eqnarray}
\ddot{\sigma}+\dot{\sigma}^2\left(1-
\frac{\sigma^2}{2Nm_P^2}\right)^{-1}
\frac{\sigma}{Nm_P^2}=\frac{V'(\sigma)}
{3H^2}H'(\sigma)\dot{\sigma}\left(1-
\frac{\sigma^2}{2Nm_P^2}\right)^{2}
\nonumber\\
-\frac{V''(\sigma)}
{3H}\dot{\sigma}\left(1-\frac{\sigma^2}{2Nm_P^2}
\right)^{2}+\frac{V'(\sigma)}{3H}\dot{\sigma}
\left(1-\frac{\sigma^2}{2Nm_P^2}\right)
\frac{\sigma}{N m_P^2}\cdot~~
\label{sigmaddot}
\end{eqnarray}
Comparing the first two terms in the RHS of
Eq.(\ref{sigmaddot}) with $H\dot{\sigma}$, we
obtain
\begin{eqnarray}
\epsilon\simeq\frac{1}{2}m_P^2\left(
\frac{V'(\sigma)}{V_0}\right)^2\left(1-
\frac{\sigma^2}{2Nm_P^2}\right)^{2}\leq 1,
\label{epsilon}\\
\vert\eta\vert\simeq m_P^2\bigg|\frac{V''
(\sigma)}{V_0}\bigg|\left(1-\frac{\sigma^2}
{2Nm_P^2}\right)^{2}\leq 1.~~~
\label{eta}
\end{eqnarray}
The third term in the RHS of Eq.(\ref{sigmaddot}),
compared to $H\dot{\sigma}$, yields $\sqrt{2}\sigma
\epsilon^{1/2}/Nm_P\leq 1$, which is automatically
satisfied provided that Eq.(\ref{epsilon}) holds
and $\sigma\leq Nm_P/\sqrt{2}$. The latter is true
for the values of $\sigma$ which are relevant here.
We see that the slow roll parameters $\epsilon$
and $\eta$ now carry an extra factor
$(1-\sigma^2/2Nm_P^2)^2\leq 1$. This leads, in
general, to smaller $\sigma_f$'s. However, in our
case, $\sigma_f\ll\sqrt{2N}m_P$ (for $N=2$) and,
thus, this factor is practically equal to unity.
Consequently, its influence on $\sigma_f$ is
negligible.

\par
The formulas for $N_Q$ and $(\delta T/T)_Q$ are
now also modified due to the presence of the
extra factor $(1-\sigma^2/2Nm_P^2)^2$ in
Eq.(\ref{infeq}). In particular, a factor
$(1-\sigma^2/2Nm_P^2)^{-2}$ must be included in
the integrand in the RHS of
Eq.(\ref{NQ}) and a factor $(1-\sigma_Q^2/2N
m_P^2)^{-4}$ in the RHS of Eq.(\ref{anisotropy}).
We find that, for the $\sigma$'s under
consideration, these modifications have only a
small influence on $\sigma_Q$ if we use the same
input values for the free parameters as in the
global SUSY case. On the contrary,
$(\delta T/T)_Q$ increases considerably.
However, we can easily readjust the parameters
so that the COBE requirements are again met.
For instance, $(\delta T/T)_Q\simeq 6.6\times
10^{-6}$ is now obtained with $m=3.8\times
10^{15}~{\rm GeV}$ keeping $\kappa=\lambda
=3\times 10^{-2}$, $\beta=0.1$ as in global
SUSY. In this case,
$\sigma_c\simeq 2.7\times 10^{16}~{\rm GeV}$,
$\sigma_f\simeq 1.8\times 10^{17}~{\rm GeV}$ and
$\sigma_Q\simeq 1.6\times 10^{18}~{\rm GeV}$.
Also, $M\simeq 2.6\times 10^{16}~{\rm GeV}$,
$N_Q\simeq 57.5$ and $n\simeq 0.99$.

\section{Conclusions}
\label{sec:conclusion}

\par
We considered the extended SUSY PS model which
has been studied in Ref.\cite{quasi}. This model
naturally leads to a moderate violation of the
`asymptotic' Yukawa coupling unification so that,
for $\mu>0$, the predicted $b$-quark mass can
take acceptable values even with universal
boundary conditions. It is \cite{quasi} also
compatible with all the other available
phenomenological and cosmological requirements.
The model contains two new superfields in the
(15,2,2) representation of $G_{PS}$. The
electroweak doublets in one of them mix with the
electroweak doublets in the usual Higgs
representation (1,2,2), thereby violating Yukawa
unification. Also, the presence of two extra
superfields $\phi$, $\bar\phi$ in the (15,1,3)
representation is necessitated by the requirement
that the violation of Yukawa unification is
adequate.

\par
We studied hybrid inflation within this model.
The inflationary superpotential
contains only renormalizable terms. In
particular, the fields $\phi$, $\bar\phi$ lead
to three new renormalizable terms which are
added to the standard (renormalizable)
superpotential for hybrid inflation. We showed
that the resulting potential possesses a
`shifted' classically flat direction which can
serve as inflationary path. We analyzed the
spectrum of the model on this path and
constructed the one-loop radiative corrections
to the potential. These corrections generate
a slope along this path which can drive the
system towards the SUSY vacuum.

\par
We find that the COBE constraint on the
quadrupole anisotropy of the CMBR can be easily
satisfied with natural values of the relevant
parameters of the model. The slow roll conditions
are violated well before the instability point of
the `shifted' path is reached and, thus,
inflation terminates smoothly. The system then
quickly approaches the critical point and, after
reaching it, enters into a `waterfall' regime
during which it falls towards the SUSY vacuum and
oscillates about it. Note that $G_{PS}$ is broken
to $G_S$ already on the `shifted' path and, thus,
there is no monopole production at the `waterfall'.

\par
As it turns out, the relevant part of inflation
occurs at values of the inflaton field which are
quite close to the `reduced' Planck scale. We,
thus, cannot ignore the SUGRA corrections which
can easily invalidate inflation by generating an
inflaton mass of the order of the Hubble constant.
To avoid this disaster, we employ the mechanism
of Ref.\cite{panag} which leads to an exact
cancellation of the inflaton mass on the
inflationary path. This mechanism relies on a
specific K\"{a}hler potential and an extra
gauge singlet with a superheavy VEV via D-terms.
We show that this mechanism readily applies to
our case. The COBE constraint can again be met
by readjusting the input values of the free
parameters which were obtained with global SUSY.

\acknowledgments
This work was supported by European Union under
the RTN contracts HPRN-CT-2000-00148 and
HPRN-CT-2000-00152. One of us (S. K.) was
supported by PPARC.

\end{document}